\documentclass[10pt,twocolumn,english,aps,manuscript,aps,showpacs,showkeys,superscriptaddress,citeautoscript]{revtex4}
\usepackage[T1]{fontenc}
\usepackage[latin9]{inputenc}
\usepackage{color}
\usepackage{amsmath}
\usepackage{amssymb}
\usepackage{graphicx}
\usepackage{esint}

\makeatletter
\@ifundefined{textcolor}{}
{%
 \definecolor{BLACK}{gray}{0}
 \definecolor{WHITE}{gray}{1}
 \definecolor{RED}{rgb}{1,0,0}
 \definecolor{GREEN}{rgb}{0,1,0}
 \definecolor{BLUE}{rgb}{0,0,1}
 \definecolor{CYAN}{cmyk}{1,0,0,0}
 \definecolor{MAGENTA}{cmyk}{0,1,0,0}
 \definecolor{YELLOW}{cmyk}{0,0,1,0}
}

\usepackage{babel}

\makeatother

\usepackage{babel}
\begin{document}

\title{Entangled state teleportation through a couple of quantum channels
composed of $XXZ$ dimers in an Ising-$XXZ$ diamond chain }

\author{M. Rojas}

\affiliation{Departamento de Física, Universidade Federal de Lavras, CP 3037,
37200-000, Lavras-MG, Brazil}

\author{S. M. de Souza}

\affiliation{Departamento de Física, Universidade Federal de Lavras, CP 3037,
37200-000, Lavras-MG, Brazil}

\author{Onofre Rojas}

\affiliation{Departamento de Física, Universidade Federal de Lavras, CP 3037,
37200-000, Lavras-MG, Brazil}

\affiliation{ICTP, Strada Costiera 11, I-34151 Trieste, Italy}
\begin{abstract}
The quantum teleportation plays an important role in quantum information
process, in this sense, the quantum entanglement properties involving
an infinite chain structure is quite remarkable because real materials
could be well represented by an infinite chain. We study the teleportation
of an entangled state through a couple of quantum channels, composed
by Heisenberg dimers in an infinite Ising-Heisenberg diamond chain,
\textcolor{black}{the couple of chains are considered sufciently far
away from each other to be ignored the any interaction between them}
. To teleporting a couple of qubits through the quantum channel, we
need to find the average density operator for Heisenberg spin dimers,
which will be used as quantum channels. Assuming the input state as
a pure state, we can apply the concept of fidelity as a useful measurement
of teleportation performance of a quantum channel. Using the standard
teleportation protocol, we have derived an analytical expression for
the output concurrence, fidelity, and average fidelity. We study in
detail the effects of coupling parameters, external magnetic field
and temperature dependence of quantum teleportation. Finally, we explore
the relations between entanglement of the quantum channel, the output
entanglement and the average fidelity of the system. Through a kind
of phase diagram as a function of Ising-Heisenberg diamond chain model
parameters, we illustrate where the quantum teleportation will succeed
and a region where the quantum teleportation could fail. 
\end{abstract}


\maketitle

\section{Introduction}

The nonlocal quantum correlation property is one of the most wonderful
types of correlation that can be shared only among quantum systems
\cite{AmicoHorod}. In recent years, many efforts have been devoted
to characterizing qualitatively and quantitatively the entanglement
properties of condensed matter systems, and has been regarded as an
essential physical resource for quantum computation and quantum communication.
In this sense, it is relevant to study the entanglement of solid state
systems such as spin chains \cite{qubit-Heisnb}. The Heisenberg spin
chain is one of the simplest quantum systems, which could exhibits
the entanglement; due to the Heisenberg exchange interaction is not
localized in the spin system.

On the other hand, in the quantum information process, the quantum
teleportation plays an important role. Since, the seminal work of
the quantum teleportation originally proposed by Bennett \cite{Bennet},
has received extensive investigations both theoretically \cite{pope}
and experimentally \cite{Bouw} in the past few years. The quantum
teleportation is a fascinating phenomenon based on a weird nonlocal
quantum property. Many schemes were proposed for teleportation, so
the system based on Heisenberg spin chain can serve as an efficient
communication channel for quantum teleportation \cite{Yeo} and reference
therein, the state of two qubits Heisenberg model has been considered
as a quantum channel in the presence and absence of an external magnetic
field. Furthermore, the quantum correlations and teleportation through
the Heisenberg $XX$ spin chain have been discussed in \cite{wan}.

It is interesting to consider the quantum antiferromagnetic Heisenberg
model on a generalized diamond chain, because this model describes
real materials such as $\mathrm{Cu_{3}(CO_{3})_{2}(OH)_{2}}$, known
as natural azurite \cite{kikuchi05}. One can associate this compound
with the model of the spin system composed of geometrically frustrated
spin on a diamond chain. In this sense, in the last decade, several
Ising-Heisenberg diamond chain structures have been discussed. There
are several approximate methods applied in Heisenberg model to explain
the experimental measurements in the natural mineral azurite \cite{Lau}.
Recently, Honecker et al. \cite{honecker} studied the dynamic and
thermodynamic properties for this model. Furthermore, thermodynamics
of the Ising-Heisenberg model on a diamond-like chain was also widely
discussed in the references \cite{Cano,orojas,valverde,lis-1}. 

Recently, in references \cite{mrojas,mrojas-1}, was investigated
the thermal entanglement in some exactly solvable infinite Ising-Heisenberg
diamond chain. Later, was also calculated the entanglement for the
hybrid diamond chain with Ising spins and electrons mobile \cite{mrojas2}.
Inspired by this works, was investigated the quantum teleportation
of two qubits in an arbitrary pure entangled state via two infinite
Ising-$XXZ$ diamond chain as quantum channel in thermal equilibrium,
studying quantities such as the output entanglement, fidelity and
average fidelity of teleportation. 

The paper is organized as follows: In Sec. II, we present the Ising-$XXZ$
model on a diamond chain. Subsequently in Sec. III, we present a brief
review of the exact solution of the model via the transfer-matrix
approach and its dimer reduced density operator. In Sec. IV, we study
analytically and numerically the fidelity, average fidelity, the concurrence
of teleported state or output state, and quantum channel concurrence.
Finally in Sec. V, is presented our conclusion.

\section{Quantum channel $XXZ$ dimer in an Ising-$XXZ$ diamond chain}

\begin{figure}
\includegraphics[scale=0.5]{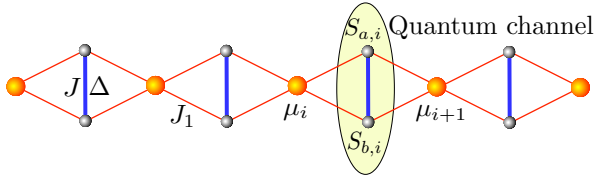}\caption{\label{fig:diamond}(Color online) Schematic representation of Ising-$XXZ$
diamond chain. The blue line represents the bipartite quantum coupling
which serves as a quantum channel. The red line corresponds to Ising
spin couplings.}
\end{figure}

We consider as a quantum channel the Ising-$XXZ$ model with nodal
Ising spins and anisotropic Heisenberg spins on a diamond-like chain
in the presence of an external magnetic field, schematically illustrated
in Fig. 1. Thus, the corresponding Hamiltonian operator can be expressed
as follow
\begin{align}
\mathcal{H}= & \sum_{i=1}^{N}[J\left(\boldsymbol{S}_{a,i},\boldsymbol{S}_{b,i}\right)_{\Delta}+J_{1}\left(S_{a,i}^{z}+S_{b,i}^{z}\right)\left(\mu_{i}+\mu_{i+1}\right)+\nonumber \\
 & -h\left(S_{a,i}^{z}+S_{b,i}^{z}\right)-\frac{h}{2}\left(\mu_{i}+\mu_{i+1}\right)],\label{eq:Hamil-1}
\end{align}
where $\left(\boldsymbol{S}_{a,i},\boldsymbol{S}_{b,i}\right)_{\Delta}=S_{a,i}^{x}S_{b,i}^{x}+S_{a,i}^{y}S_{b,i}^{y}+\Delta S_{a,i}^{z}S_{b,i}^{z}$
corresponds to the interstitial anisotropic Heisenberg spins coupling
$J$ and $\Delta$, whereas the nodal-interstitial spins $\mu_{i}$
are representing the Ising-type exchanges $J_{1}$. The system is
under longitudinal external magnetic field $h$ acting on Heisenberg
spins and Ising spins. 

The quantum Heisenberg spin coupling can be expressed using matrix
notation in the standard basis $\{|\uparrow\uparrow\rangle,|\uparrow\downarrow\rangle,|\downarrow\uparrow\rangle,|\downarrow\downarrow\rangle\}$,
where $|\uparrow\rangle$ and $|\downarrow\rangle$ denote the spin-up
and spin-down states, respectively. Thus, we have
\begin{equation}
\left(\boldsymbol{S}_{a,i},\boldsymbol{S}_{b,i}\right)_{\Delta}=\left[\begin{array}{cccc}
\frac{\Delta}{4} & 0 & 0 & 0\\
0 & -\frac{\Delta}{4} & \frac{1}{2} & 0\\
0 & \frac{1}{2} & -\frac{\Delta}{4} & 0\\
0 & 0 & 0 & \frac{\Delta}{4}
\end{array}\right],
\end{equation}
and
\begin{equation}
S_{a,i}^{z}+S_{b,i}^{z}=\left[\begin{array}{cccc}
1 & 0 & 0 & 0\\
0 & 0 & 0 & 0\\
0 & 0 & 0 & 0\\
0 & 0 & 0 & -1
\end{array}\right].
\end{equation}

Thus, we obtain the following eigenvalues after the diagonalization
of Heisenberg spins dimer (sites $a$ and $b$), and assuming fixed
values for $\mu_{i}$ and $\mu_{i+1}$, we have
\begin{align}
\mathcal{E}_{1}(\mu_{i},\mu_{i+1})= & \frac{J\Delta}{4}+\left(J_{1}-\frac{h}{2}\right)\left(\mu_{i}+\mu_{i+1}\right)-h,\nonumber \\
\mathcal{E}_{2}(\mu_{i},\mu_{i+1})= & \frac{J}{2}-\frac{J\Delta}{4}-\frac{h}{2}\left(\mu_{i}+\mu_{i+1}\right),\nonumber \\
\mathcal{E}_{3}(\mu_{i},\mu_{i+1})= & -\frac{J}{2}-\frac{J\Delta}{4}-\frac{h}{2}\left(\mu_{i}+\mu_{i+1}\right),\nonumber \\
\mathcal{E}_{4}(\mu_{i},\mu_{i+1})= & \frac{J\Delta}{4}-\left(J_{1}+\frac{h}{2}\right)\left(\mu_{i}+\mu_{i+1}\right)+h.\label{eq:eigenvals}
\end{align}
Where their corresponding eigenstates are obtained using standard
basis respectively by
\begin{align}
|\varphi_{1}\rangle= & |\uparrow\uparrow\rangle,\\
|\varphi_{2}\rangle= & \frac{1}{\sqrt{2}}\left(|\uparrow\downarrow\rangle+|\downarrow\uparrow\rangle\right),\label{eq:phi2}\\
|\varphi_{3}\rangle= & \frac{1}{\sqrt{2}}\left(|\uparrow\downarrow\rangle-|\downarrow\uparrow\rangle\right),\label{eq:phi3}\\
|\varphi_{4}\rangle= & |\downarrow\downarrow\rangle.
\end{align}

\subsection{Quantum channel density operator}

The Ising-$XXZ$ diamond chain was recently studied \cite{mrojas},
through a decoration transformation \cite{phys-A-09-1,strecka pla-1}
and a transfer-matrix approach \cite{baxter-book-1}. 

Guided by that solution, the local density operator for dimer operator
(site $a$ and $b$) bonded by Ising particles $\mu$ and $\mu'$,
can be expressed by 
\begin{equation}
\varrho(\mu,\mu')=\sum_{i=1}^{4}\mathrm{e}^{-\beta\varepsilon_{i}(\mu,\mu')}|\varphi_{i}\rangle\langle\varphi_{i}|.
\end{equation}

Here, $\beta=1/(k_{B}T)$, $k_{B}$ is being the Boltzmann`s, constant
and $T$ being the absolute temperature.

Thus, the density operator in the natural basis can be expressed by
\begin{equation}
\varrho=\left[\begin{array}{cccc}
\varrho_{1,1} & 0 & 0 & 0\\
0 & \varrho_{2,2} & \varrho_{2,3} & 0\\
0 & \varrho_{3,2} & \varrho_{3,3} & 0\\
0 & 0 & 0 & \varrho_{4,4}
\end{array}\right],
\end{equation}
where the elements of the two qubits operator are
\begin{align}
\varrho_{1,1}(\mu,\mu')= & \mathrm{e}^{-\beta\varepsilon_{1}(\mu,\mu')},\nonumber \\
\varrho_{2,2}(\mu,\mu')= & \frac{1}{2}\left(\mathrm{e}^{-\beta\varepsilon_{2}(\mu,\mu')}+\mathrm{e}^{-\beta\varepsilon_{3}(\mu,\mu')}\right),\nonumber \\
\varrho_{2,3}(\mu,\mu')= & \frac{1}{2}\left(\mathrm{e}^{-\beta\varepsilon_{2}(\mu,\mu')}-\mathrm{e}^{-\beta\varepsilon_{3}(\mu,\mu')}\right),\nonumber \\
\varrho_{4,4}(\mu,\mu')= & \mathrm{e}^{-\beta\varepsilon_{4}(\mu,\mu')}.
\end{align}

\subsection{Average density operator of Heisenberg spin dimer}

Following the result obtained in reference \cite{mrojas}, the average
density operator of Heisenberg spin dimer, is given by

\begin{equation}
\rho_{ch}=\left[\begin{array}{cccc}
\rho_{1,1} & 0 & 0 & 0\\
0 & \rho_{2,2} & \rho_{2,3} & 0\\
0 & \rho_{3,2} & \rho_{3,3} & 0\\
0 & 0 & 0 & \rho_{4,4}
\end{array}\right],\label{eq:rho-mat}
\end{equation}
where elements of quantum channel density operator, in the thermodynamic
limit, are expressed
\begin{align}
\rho_{i,j}= & \frac{1}{\Lambda_{+}}\left\{ \tfrac{\varrho_{i,j}(\tfrac{1}{2},\tfrac{1}{2})+\varrho_{i,j}(-\tfrac{1}{2},-\tfrac{1}{2})}{2}+\tfrac{2\varrho_{i,j}(\tfrac{1}{2},-\tfrac{1}{2})w_{+-}}{\sqrt{\left(w_{++}-w_{--}\right)^{2}+4w_{+-}^{2}}}\right.\nonumber \\
 & \left.+\tfrac{\left(\varrho_{i,j}(\tfrac{1}{2},\tfrac{1}{2})-\varrho_{i,j}(-\tfrac{1}{2},-\tfrac{1}{2})\right)\left(w_{++}-w_{--}\right)}{2\sqrt{\left(w_{++}-w_{--}\right)^{2}+4w_{+-}^{2}}}\right\} .\label{eq:rho-elem-1}
\end{align}
where the Boltzmann factor is given by
\begin{equation}
w(\mu,\mu')=\mathrm{tr}_{ab}\left(\varrho(\mu,\mu')\right)=\sum_{i=1}^{4}\mathrm{e}^{-\beta\varepsilon_{i}(\mu,\mu')}.\label{eq:w-def}
\end{equation}
Defining conveniently $w_{\pm\pm}\equiv w(\pm\frac{1}{2},\pm\frac{1}{2})$
and $w_{+-}\equiv w(\frac{1}{2},-\frac{1}{2})$. Whereas, $\Lambda_{+}$
is the largest eigenvalue
\begin{equation}
\Lambda_{+}=\tfrac{w_{++}+w_{--}+\sqrt{\left(w_{++}-w_{--}\right)^{2}+4w_{+-}^{2}}}{2}.
\end{equation}

The average density operator \eqref{eq:rho-mat} of Heisenberg spin
dimer, will be useful to study quantum teleportation through the Heisenberg
dimers.

\section{Entangled state teleportation }

\begin{figure}

\includegraphics[scale=0.45]{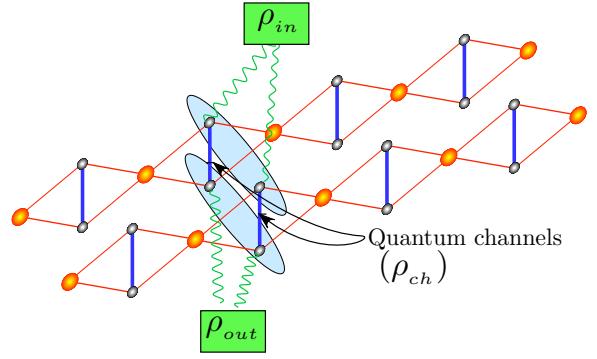}\caption{\label{fig:teleport} Schematic representation for teleportation of
input state $\rho_{in}$, through a couple of independent Heisenberg
dimers (blue lines) in an Ising-Heisenberg diamond chain, and the
teleported output state is denoted by $\rho_{out}$.}
\end{figure}

In this section, we study the quantum teleportation using as quantum
channel a couple of Heisenberg dimer in the Ising-Heisenberg diamond
chain discussed above, considering the standard teleportation protocol
\cite{Bennet}. \textcolor{black}{The couple of quantum channels are
considered sufficiently far away from each other, thus we ignore any
possible coupling between each diamond chains.} The standard teleportation
of two qubits through a mixed entangled state; can be viewed as a
generalized depolarizing channel \cite{bowen,horo}. The input state
$\rho_{in}=|\psi_{in}\rangle\langle\psi_{in}|$ depicted in figure
\ref{fig:teleport} is destroyed and its output state $\rho_{out}$
appears on another side of the diamond chain after applying a local
measurement in the form of linear operators.

Let us consider an initial state of two qubits in the natural basis
of qubits $\{|0\rangle,|1\rangle\}$, note that this basis cannot
be confused with Heisenberg dimer natural basis, defined previously.
Thus, the initial unknown pure state is expressed by
\begin{equation}
\begin{array}{ccc}
\mid\psi_{in}\rangle & = & \cos(\frac{\theta}{2})|10\rangle+{\rm e}^{i\phi}\sin(\frac{\theta}{2})|01\rangle,\end{array}
\end{equation}
 where $0\leqslant\theta\leqslant\pi$ and $0\leqslant\phi\leqslant2\pi$.
Here, $\theta$ describes an arbitrary state and $\phi$ is the corresponding
phase of this state. 

Surely, the initial state $\rho_{in}$ concurrence, can be obtained
easily, which becomes 
\begin{equation}
\mathcal{C}_{in}=2\left|e^{i\phi}\sin\left(\tfrac{\theta}{2}\right)\cos\left(\tfrac{\theta}{2}\right)\right|=|\sin\left(\theta\right)|.\label{eq:Cin}
\end{equation}

To study the output state $\rho_{out}$, we need to use the well known
Bell states given by

\begin{alignat}{1}
|\Phi^{\pm}\rangle= & \tfrac{1}{\sqrt{2}}\left(|00\rangle\pm|11\rangle\right),\\
|\Psi^{\pm}\rangle= & \tfrac{1}{\sqrt{2}}\left(|01\rangle\pm|10\rangle\right),
\end{alignat}
with these states, we can construct the projection operator for each
Bell states: $E^{0}=|\Psi^{-}\rangle\langle\Psi^{-}|$, $E^{1}=|\Phi^{-}\rangle\langle\Phi^{-}|$,
$E^{2}=|\Phi^{+}\rangle\langle\Phi^{+}|$ and $E^{3}=|\Psi^{+}\rangle\langle\Psi^{+}|$.
Besides, the probability to find a Bell state is given by $p_{i}={\rm tr}\left[E^{i}\rho_{ch}\right]$.
Obviously, we can verify that $\underset{i}{\sum}p_{i}=1$.

Therefore, it is possible to express the output state $|\rho_{out}\rangle$
using the density operator \cite{bowen} defined by, 
\begin{equation}
\rho_{out}=\underset{i,j=\{0,x,y,z\}}{\sum}p_{i}p_{j}(\sigma_{i}\otimes\sigma_{j})\rho_{in}(\sigma_{i}\otimes\sigma_{j}),
\end{equation}
 where $\sigma_{0}$ is the identity matrix and $\sigma_{\alpha}$
$(\alpha=x,y,z)$ are the three components of the Pauli matrices. 

Now, let us express the elements of density operator $\rho_{out}$,
which has the following structure
\begin{equation}
\rho_{out}=\left[\begin{array}{cccc}
\alpha & 0 & 0 & 0\\
0 & a & b & 0\\
0 & b^{*} & d & 0\\
0 & 0 & 0 & \alpha
\end{array}\right].
\end{equation}
Where the elements of average density operator can be expressed by,
\begin{align}
\alpha=\: & 2\rho_{2,2}\left(\rho_{1,1}+\rho_{4,4}\right),\nonumber \\
a=\: & \left(\rho_{1,1}+\rho_{4,4}\right)^{2}\cos^{2}\left(\tfrac{\theta}{2}\right)+4\rho_{2,2}^{2}\sin^{2}\left(\tfrac{\theta}{2}\right),\nonumber \\
b=\: & 2e^{i\phi}\rho_{2,3}^{2}\sin\theta,\nonumber \\
d=\: & 4\rho_{2,2}^{2}\cos^{2}\left(\tfrac{\theta}{2}\right)+\left(\rho_{1,1}+\rho_{4,4}\right)^{2}\sin^{2}\left(\tfrac{\theta}{2}\right).
\end{align}

To describe the thermal entanglement of the output state \textcolor{black}{$\rho_{out}$,}
we use the concurrence defined by Wootters \cite{wootters,hill},
which is given by
\begin{equation}
\mathcal{C}_{out}=\mathrm{max}\{\sqrt{\lambda_{1}}-\sqrt{\lambda_{2}}-\sqrt{\lambda_{3}}-\sqrt{\lambda_{4}},0\},\label{eq:Cdf}
\end{equation}
assuming $\lambda_{i}$ are the eigenvalues in decreasing order of
the matrix
\begin{equation}
R_{out}=\rho_{out}\left(\sigma^{y}\otimes\sigma^{y}\right)\rho_{out}^{*}\left(\sigma^{y}\otimes\sigma^{y}\right),\label{eq:R}
\end{equation}
where $\rho_{out}^{*}$ denotes the complex conjugation of $\rho_{out}$.

It is easy to show, that the eigenvalues of Eq. \eqref{eq:R}, are
given by 
\begin{align}
\lambda_{1(2)}= & \left(\sqrt{ad}\pm\sqrt{bb^{*}}\right)^{2},\; & \lambda_{3(4)}=\alpha^{2},\label{eq:lambdas-out}
\end{align}
with 
\begin{alignat}{1}
ad= & \left(\tfrac{\left(\rho_{1,1}+\rho_{4,4}\right)^{2}}{2}+2\rho_{2,2}^{2}\right)^{2}-\nonumber \\
 & \left(\tfrac{\left(\rho_{1,1}+\rho_{4,4}\right)^{2}}{2}-2\rho_{2,2}^{2}\right)^{2}\left(1-\mathcal{C}_{in}^{2}\right),\\
bb^{*}= & 4\rho_{2,3}^{4}\mathcal{C}_{in}^{2}.
\end{alignat}

Using the result Eq. \eqref{eq:lambdas-out}, the concurrence of the
output state can be obtained from Eq. (\ref{eq:Cdf}), which results
in 
\begin{equation}
\mathcal{C}_{out}=2\mathrm{max}\left\{ 2\rho_{2,3}^{2}\mathcal{C}_{in}-2\mid\rho_{2,2}\mid\mid\rho_{1,1}+\rho_{4,4}\mid,0\right\} .\label{eq:conc}
\end{equation}

\begin{figure}
\includegraphics[scale=0.23]{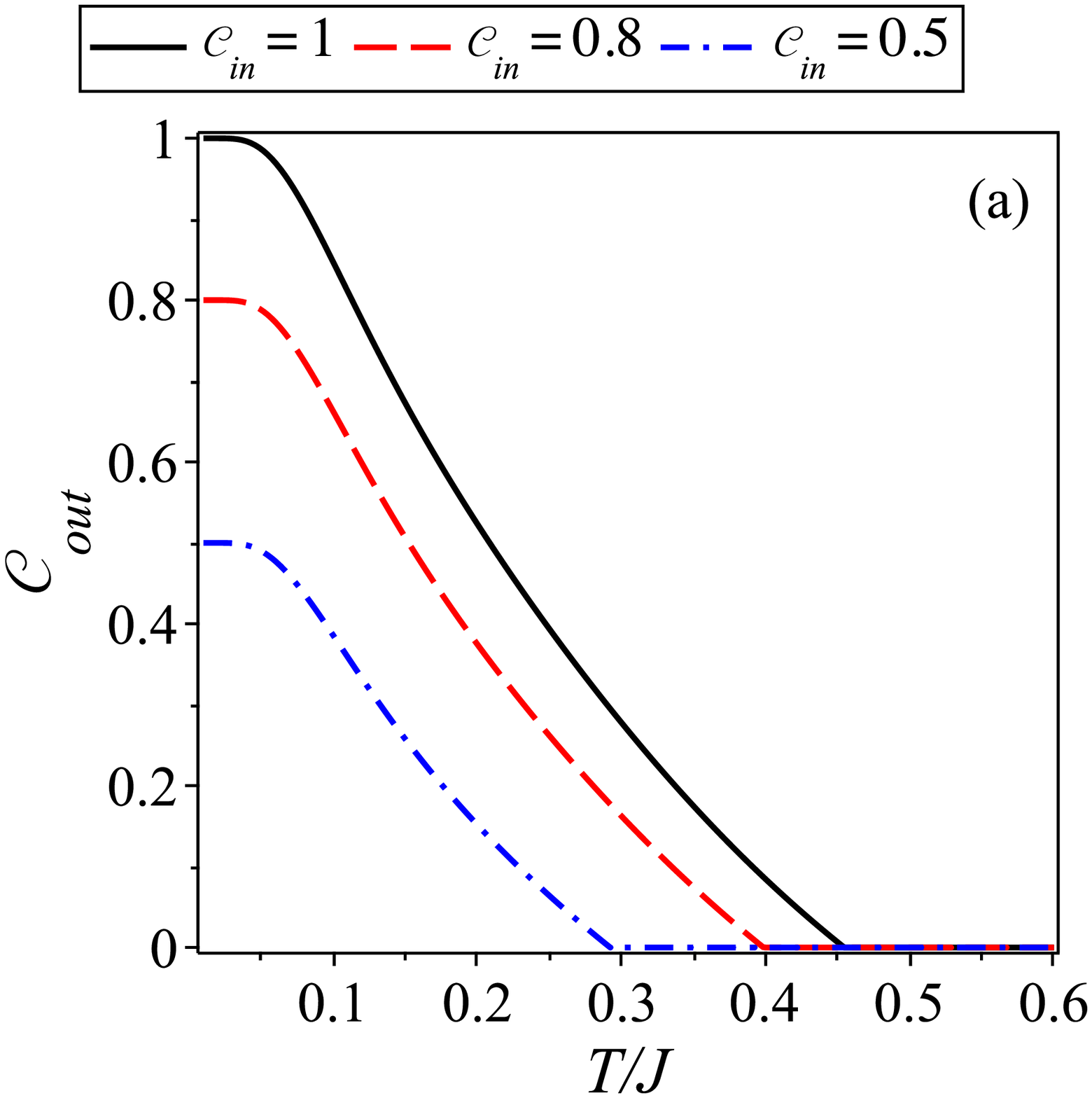}\includegraphics[scale=0.23]{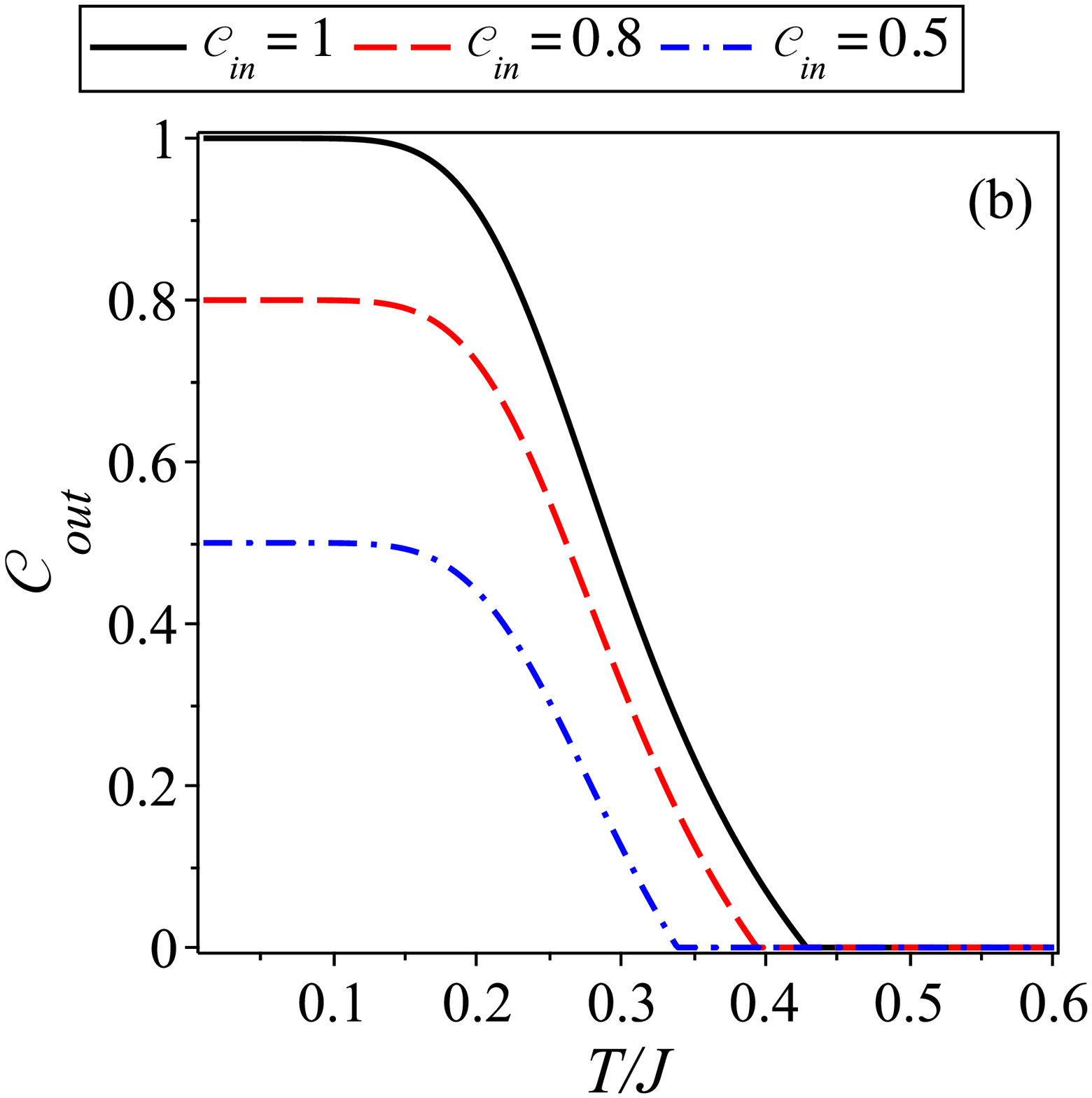}\caption{\label{fig:C-T}(Color online) Output concurrence $\mathcal{C}_{out}$
as a function of $T/J$ for $J_{1}/J=1$, $\Delta=1.5$ and different
values of $\mathcal{C}_{in}$. In (a) we display $\mathcal{C}_{out}$
in the absence of the magnetic field and in (b) we display $\mathcal{C}_{out}$
for magnetic field $h/J=1$.}
\end{figure}

In what follows, study the effects of the output entanglement as a
function of the concurrence and quantum channel condition (the Ising-Heisenberg
parameters). In Fig. \ref{fig:C-T}(a) is illustrated the output concurrence
$\mathcal{C}_{out}$ as a function of the $T/J$ for different values
of input concurrence $\mathcal{C}_{in}$, assuming null magnetic field.
We observe the output concurrence decreases as soon as the temperature
increases and for temperature higher than the threshold temperature,
the output concurrence becomes null indicating there is no entanglement
for temperature above than threshold temperature. In the low-temperature
limit, we can also observe the output concurrence and input concurrence
is closely related. Whereas, in Fig. \ref{fig:C-T}(b) is illustrated
for $h/J=1$, observing the output concurrence is enhanced in the
low temperature region due to de presence of magnetic field. Although
for higher temperature the entanglement is also destroyed.

\begin{figure}
\includegraphics[scale=0.3]{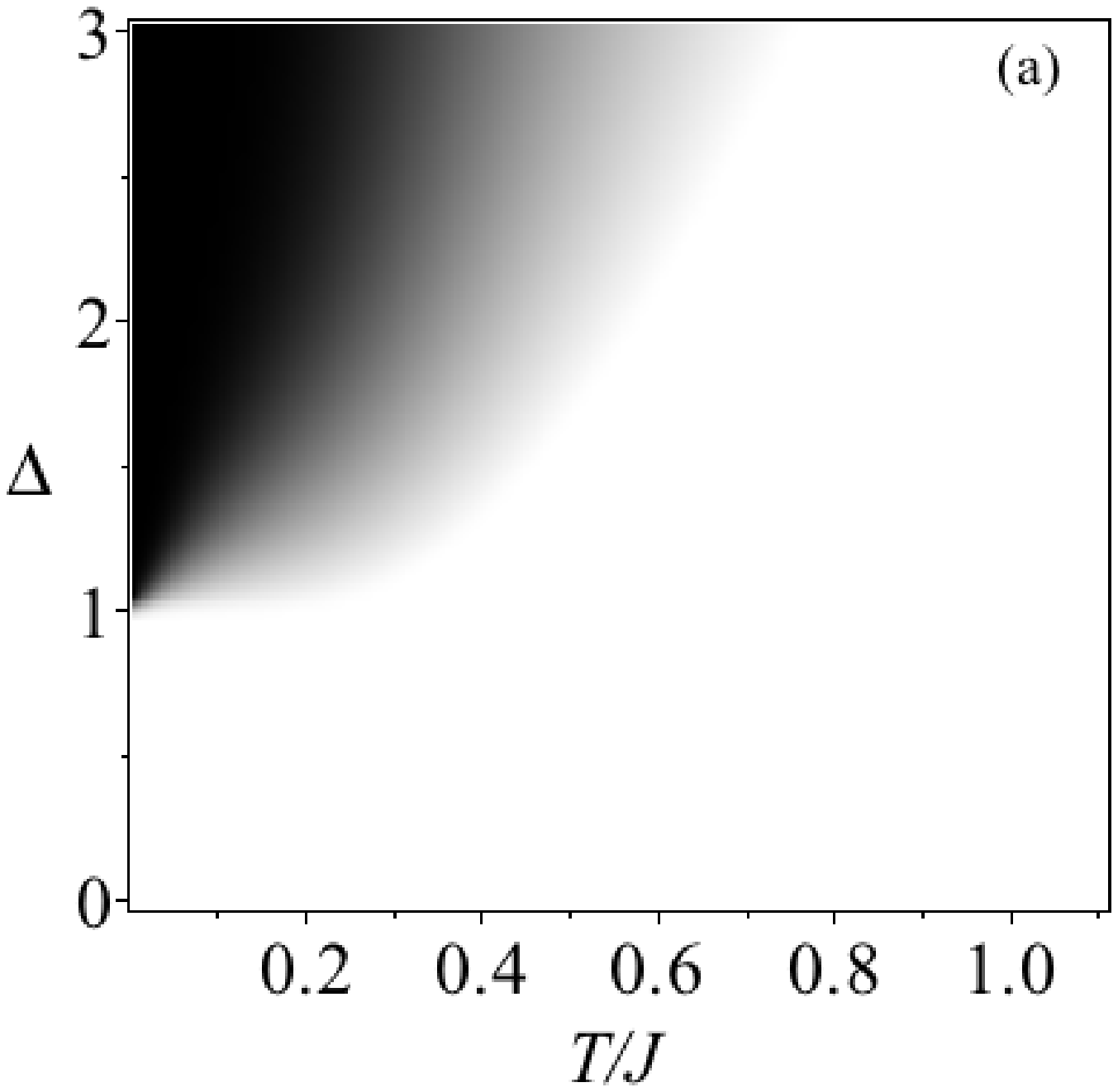}\includegraphics[scale=0.3]{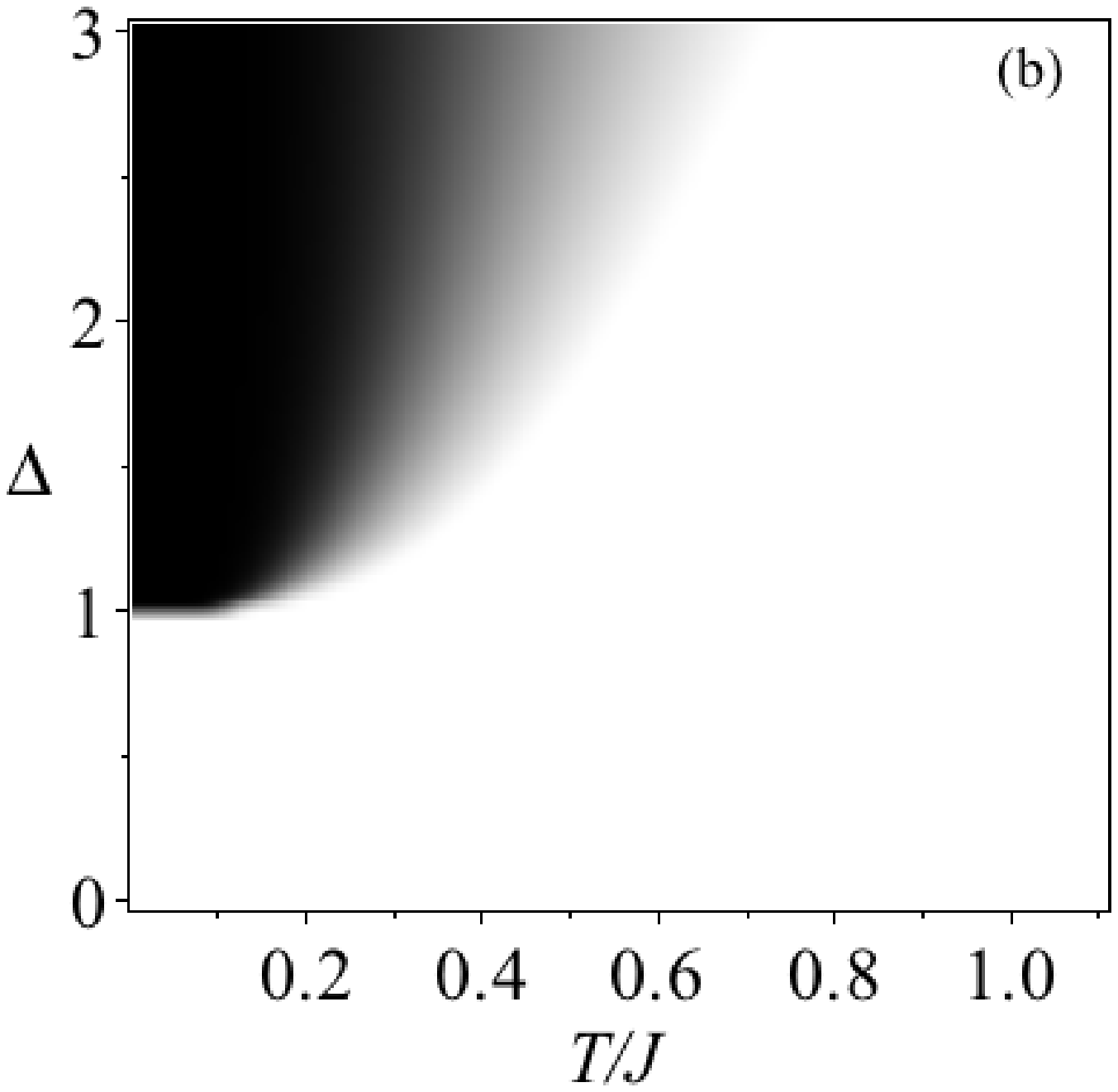}\caption{\label{fig:TvsDlt}Density plot of output concurrence $\mathcal{C}_{out}$
as a function of $T/J$ and $\Delta$. In (a) we display $\mathcal{C}_{out}$
in the absence of the magnetic field. In (b) we display $\mathcal{C}_{out}$
for $h/J=1$, the black (white) region corresponds to $\mathcal{C}_{out}=1(0)$
and by gray regions we indicate a concurrence $0<\mathcal{C}_{out}<1$.}
\end{figure}

Now we can start our discussion regarding quantum teleportation of
the entangled input state. From Eq. (\ref{eq:conc}) we can analyze
the behavior of teleported state $\rho_{out}$, for a range of parameters
assuming the state $\rho_{in}$ is maximally entangled ($\theta=\pi/2$).

In Fig. \ref{fig:TvsDlt}, we illustrate the density plot of output
concurrence $\mathcal{C}_{out}$ as a function of $T/J$ and $\Delta$
for a fixed value of $J_{1}/J$=1. The black region corresponds to
the maximum entangled region ($\mathcal{C}_{out}=1$), whereas the
white region corresponds to the unentangled region ($\mathcal{C}_{out}=0$).
The gray region means the different degrees of entanglement ($0<\mathcal{C}_{out}<1$).
In Fig. \ref{fig:TvsDlt}(a) it is shown that the model is maximally
entangled only for $\Delta\geqslant1$ in the absence of the magnetic
field, while the concurrence $\mathcal{C}_{out}$ is always null for
$\Delta<1$. The concurrence $\mathcal{C}_{out}$ becomes smaller
when increasing the temperature and as expected the entanglement vanishes
at high temperature. In Fig. \ref{fig:TvsDlt}(b), we display $\mathcal{C}_{out}$
for $h/J=1$, where the concurrence behaves similar to Fig. \ref{fig:TvsDlt}(a),
also the concurrence is enhanced up to higher temperature.

\begin{figure}
\includegraphics[scale=0.305]{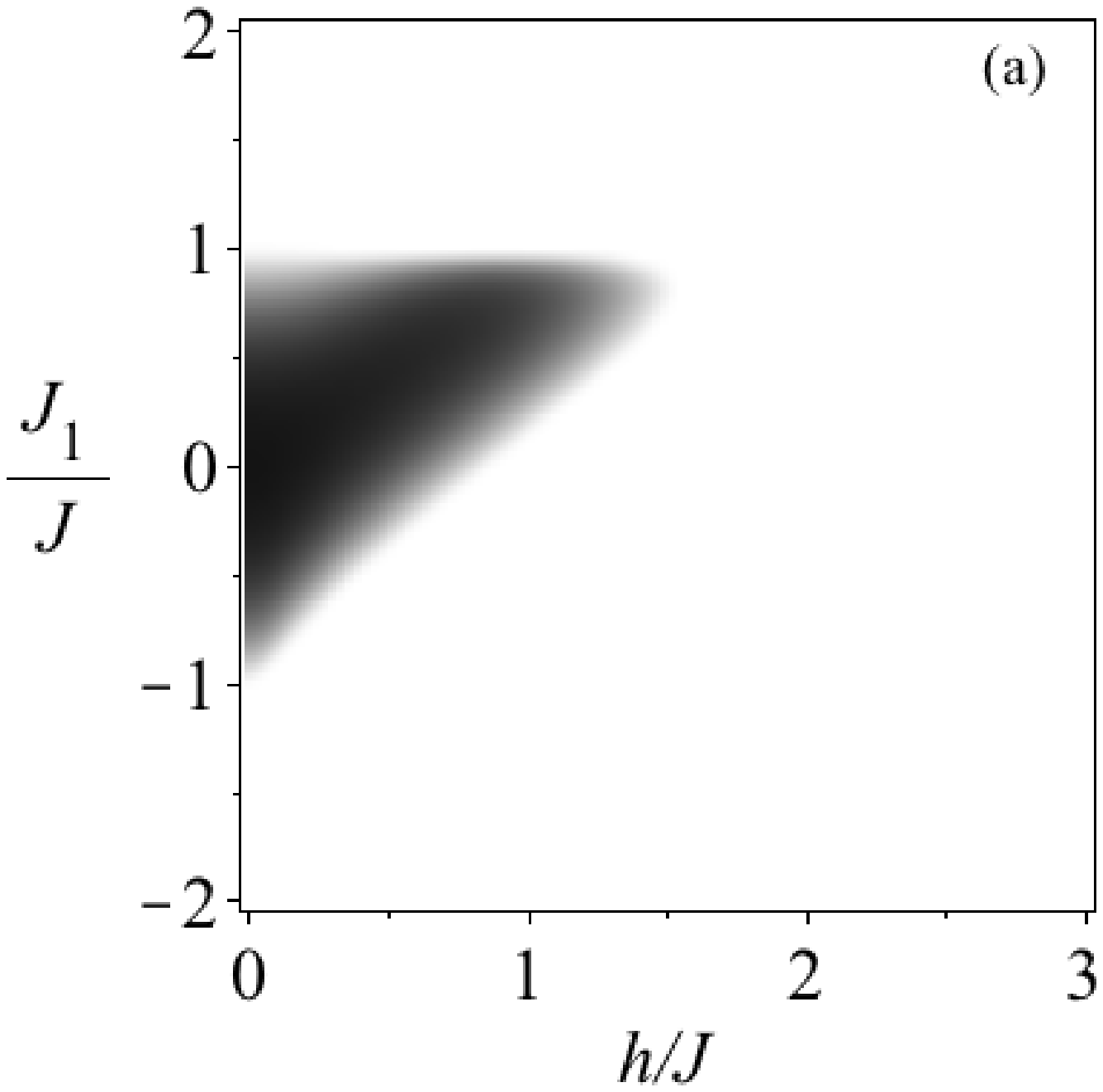}\includegraphics[scale=0.305]{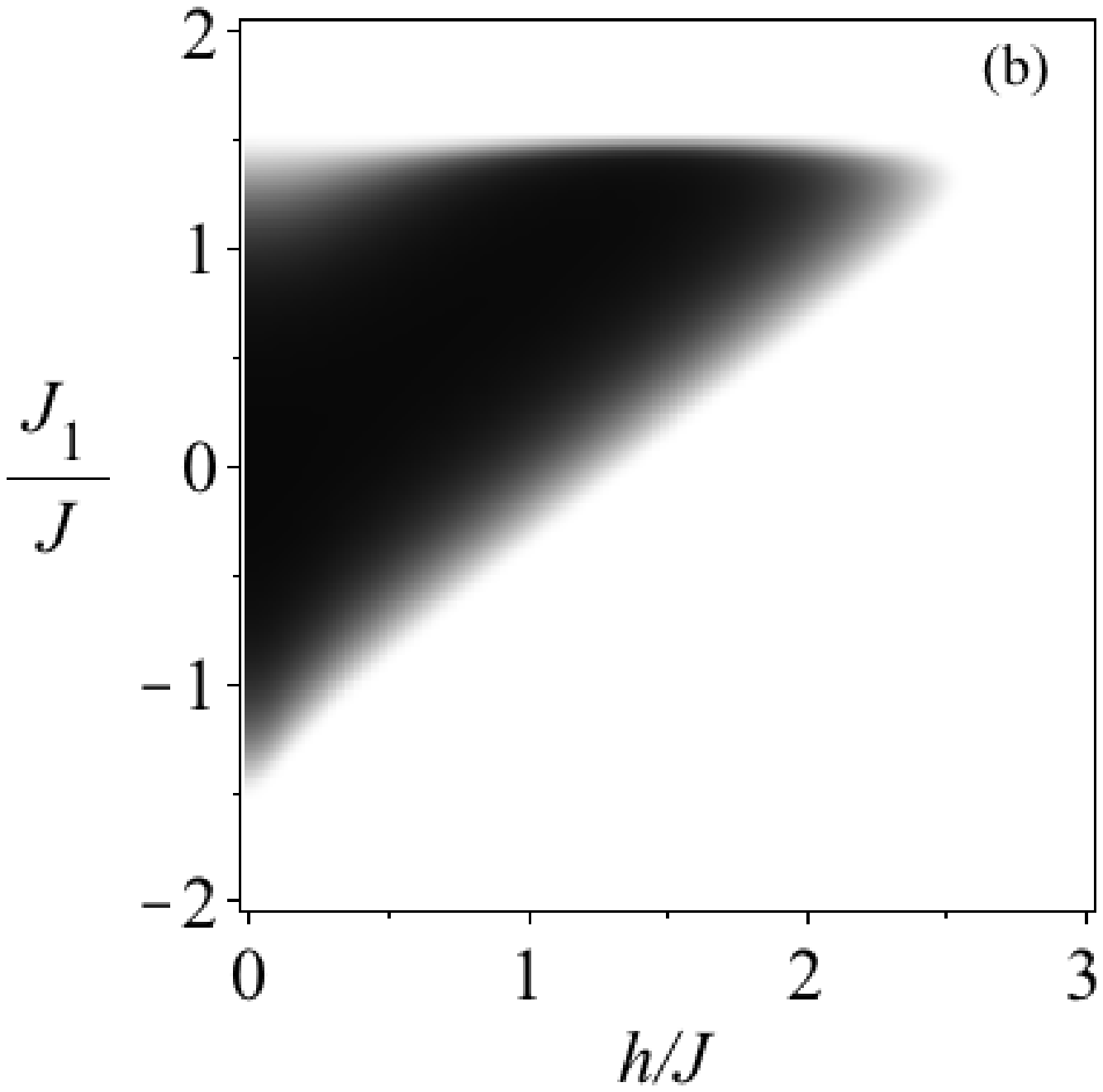}

\caption{\label{fig:hvsJ1}Density plot of output concurrence $\mathcal{C}_{out}$
as a function of $h/J$ and $J_{1}/J$ . In (a) we display of $\Delta=1$
and in (b) is for $\Delta=2$.}
\end{figure}

In Fig. \ref{fig:hvsJ1} is shown another density plot of concurrence
$\mathcal{C}_{out}$ as a function of magnetic field $h/J$ and $J_{1}/J$.
To represent the concurrence, we use the same representation as in
Fig. \ref{fig:TvsDlt}, assuming fixed value of $T/J=0.2$. In Fig.
\ref{fig:hvsJ1}(a), we display $\mathcal{C}_{out}$ for $\Delta=1$,
thus, we can illustrate that the concurrence is always less than $\mathcal{C}_{out}\lesssim0.9$.
The largest concurrence occurs for $|J_{1}/J|\lesssim1$ and magnetic
field $h/J\lesssim1.5$. Whereas, in Fig. \ref{fig:hvsJ1}(b), we
display $\mathcal{C}_{out}$ for $\Delta=2$, and we observe that,
the output concurrence is enhanced by increasing the anisotropy $\Delta$.
Therefore, the teleportation is more efficient in the environment
with a high magnetic field and strong anisotropy parameter of Ising-Heisenberg
chain (quantum channel condition).

\section{Fidelity of entangled state teleportation }

To describe the quality of the process of teleportation, it is quite
relevant to study the fidelity between $\rho_{in}$ and $\rho_{out}$
to characterizes the success of teleported state. When the input state
is a pure state, we can apply the concept of fidelity as a useful
indicator of teleportation performance of a quantum channel. The fidelity
of $\rho_{out}$ was defined\textcolor{red}{{} }\cite{Joz} as 
\begin{eqnarray}
F & = & \langle\psi_{in}|\rho_{out}|\psi_{in}\rangle,\nonumber \\
 & = & \left\{ {\rm tr}\left[\sqrt{\sqrt{\rho_{in}}\rho_{out}\sqrt{\rho_{in}}}\right]\right\} ^{2}.\label{eq:fid}
\end{eqnarray}
After a straightforward calculation, the fidelity becomes
\begin{equation}
F=\tfrac{\sin^{2}\theta}{2}\left[\left(\rho_{1,1}+\rho_{4,4}\right)^{2}+4\rho_{2,3}^{2}-4\rho_{2,2}^{2}\right]+4\rho_{2,2}^{2}.
\end{equation}

When the input state is a pure state, the efficiency of quantum communication
is characterized by the average fidelity \cite{Joz}. The average
fidelity $F_{A}$ of teleportation can be formulated as 
\begin{equation}
F_{A}=\frac{1}{4\pi}\intop_{0}^{2\pi}d\phi\intop_{0}^{\pi}F\sin\theta d\theta.
\end{equation}
After integrating the average fidelity $F_{A}$ we obtain 
\begin{equation}
F_{A}=\frac{1}{3}\left[\left(\rho_{1,1}+\rho_{4,4}\right)^{2}+4\rho_{2,3}^{2}-4\rho_{2,2}^{2}\right]+4\rho_{2,2}^{2}.\label{eq:Fa1}
\end{equation}

To transmit a quantum state $|\psi_{in}\rangle$ better than any classical
communication protocol, $F_{A}$ must be greater than $\frac{2}{3}$
which is the best fidelity in the classical world \cite{Joz}. 

To describe the behavior of the average fidelity in the standard teleportation
protocol, some plots are given below.

In Fig. \ref{fig:TvsJ1} we depict the average fidelity as a function
of $\Delta$ and $T/J$ for a fixed value of $J_{1}/J=1$. The black
region correspond to the maximum average fidelity $\left(F_{A}=1\right)$,
while the white region corresponds to $F_{A}=0$. The yellow curve
surrounding the dark region ($F_{A}>2/3$) means the region in which
the quantum teleportation will become successful, whereas the outside
means the quantum teleportation fails and becoming classical communication
region. From Fig. \ref{fig:TvsJ1}(a), we can see that when $\Delta\geqslant1$
the average fidelity is maximum in the absence of magnetic field,
while for $\Delta<1$ the average fidelity is always less than or
equal to $2/3$. In Fig. \ref{fig:TvsJ1}(b), we display $F_{A}$
for $h/J=1$, where the average fidelity behaves similar to Fig. \ref{fig:TvsJ1}(a).
However, the Fig. \ref{fig:TvsJ1}(b) clearly shows that, when $\Delta\thickapprox1$
and at temperatures less than $T/J\approx0.1$, the average fidelity
is maximum.

\begin{figure}
\includegraphics[scale=0.3]{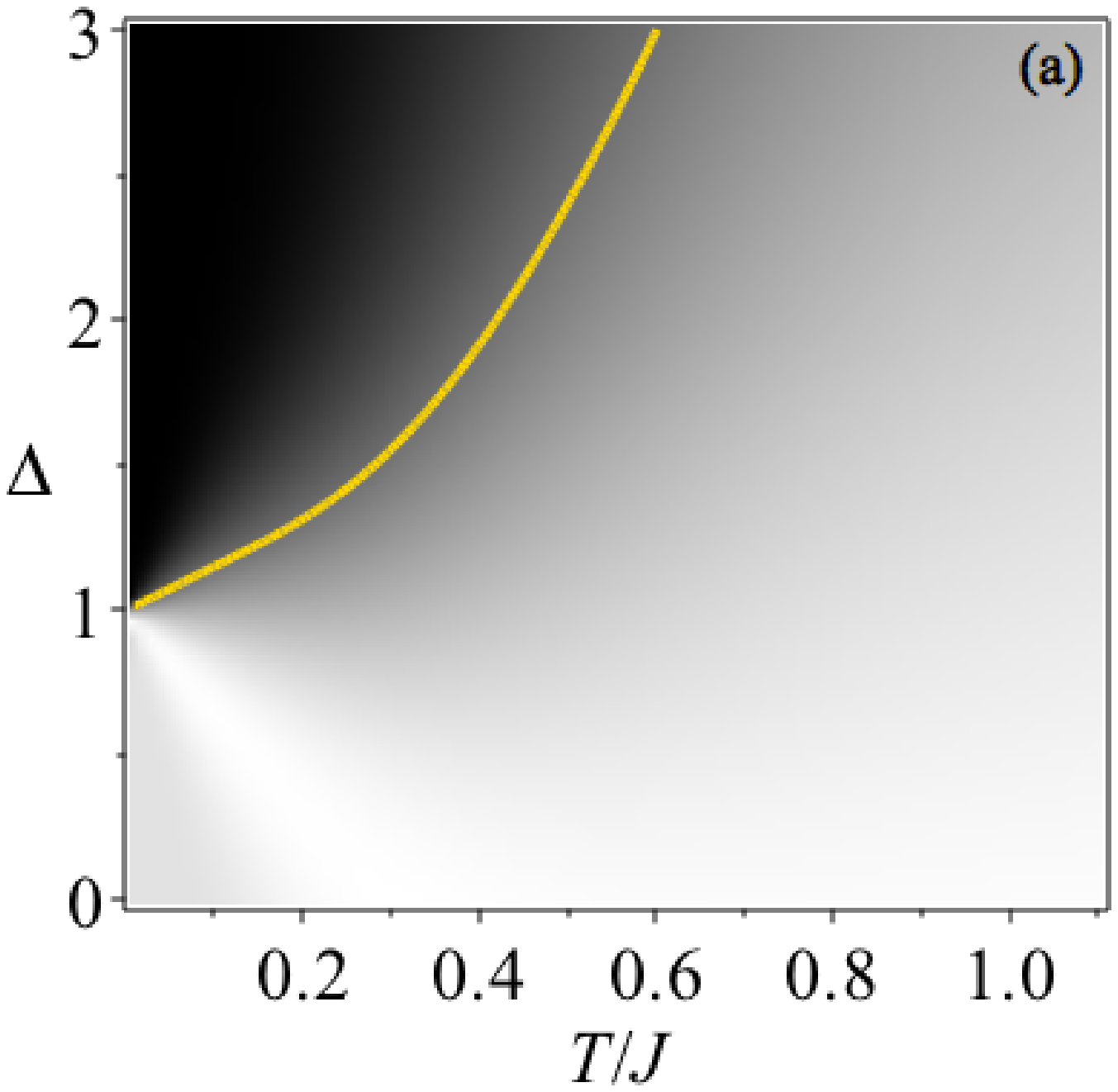}\includegraphics[scale=0.3]{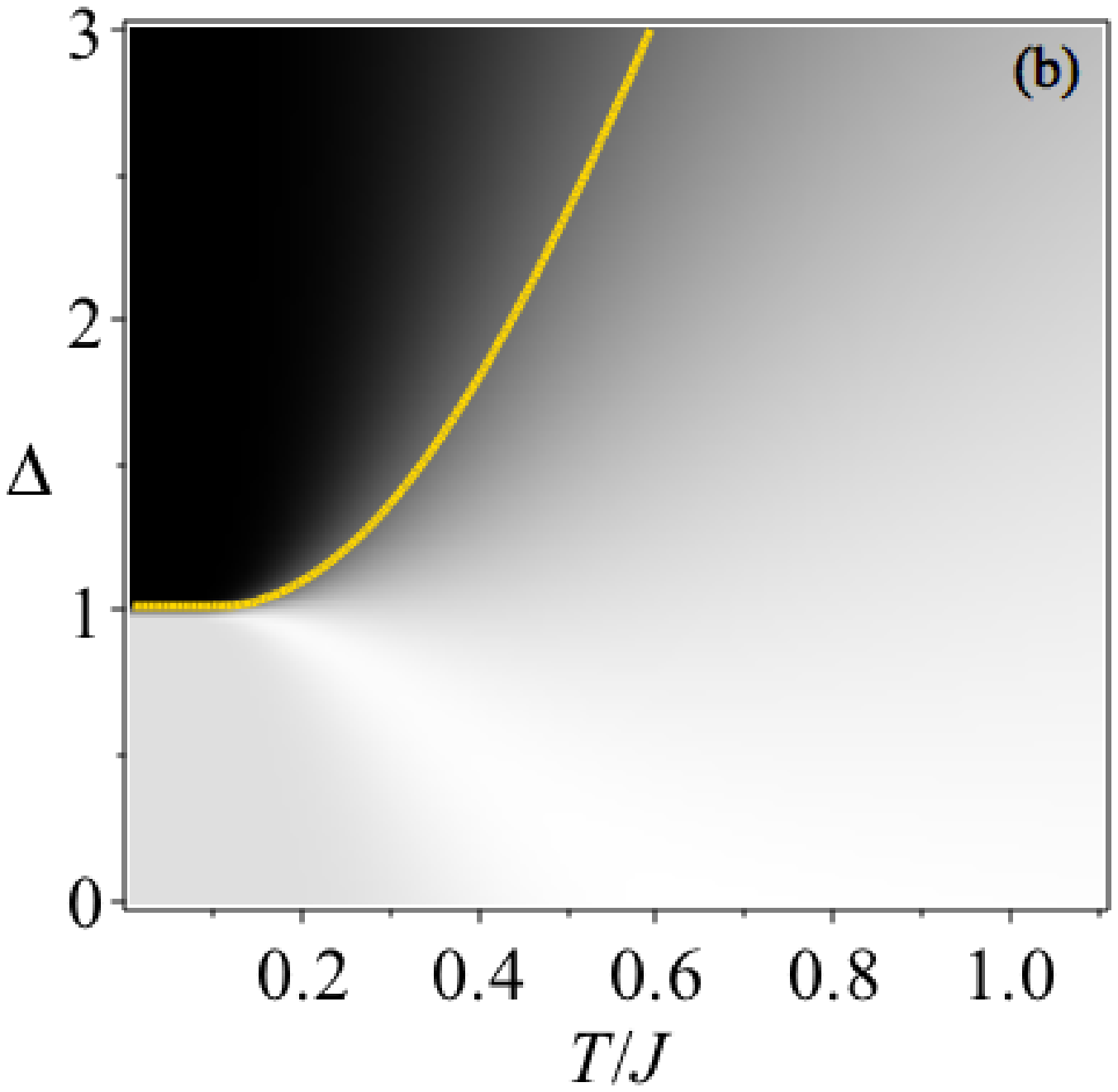}\caption{\label{fig:TvsJ1} Density plot average of fidelity $F_{A}$ as a
function of $T/J$ versus $\Delta$ for $J_{1}/J=1.0$. In (a) we
display $F_{A}$ for a $h/J=0$ and in (b) we display $F_{A}$ for
$h/J=1$. For both figures, the yellow curve is the contour for $F_{A}=2/3$.}
\end{figure}

\begin{figure}
\includegraphics[scale=0.3]{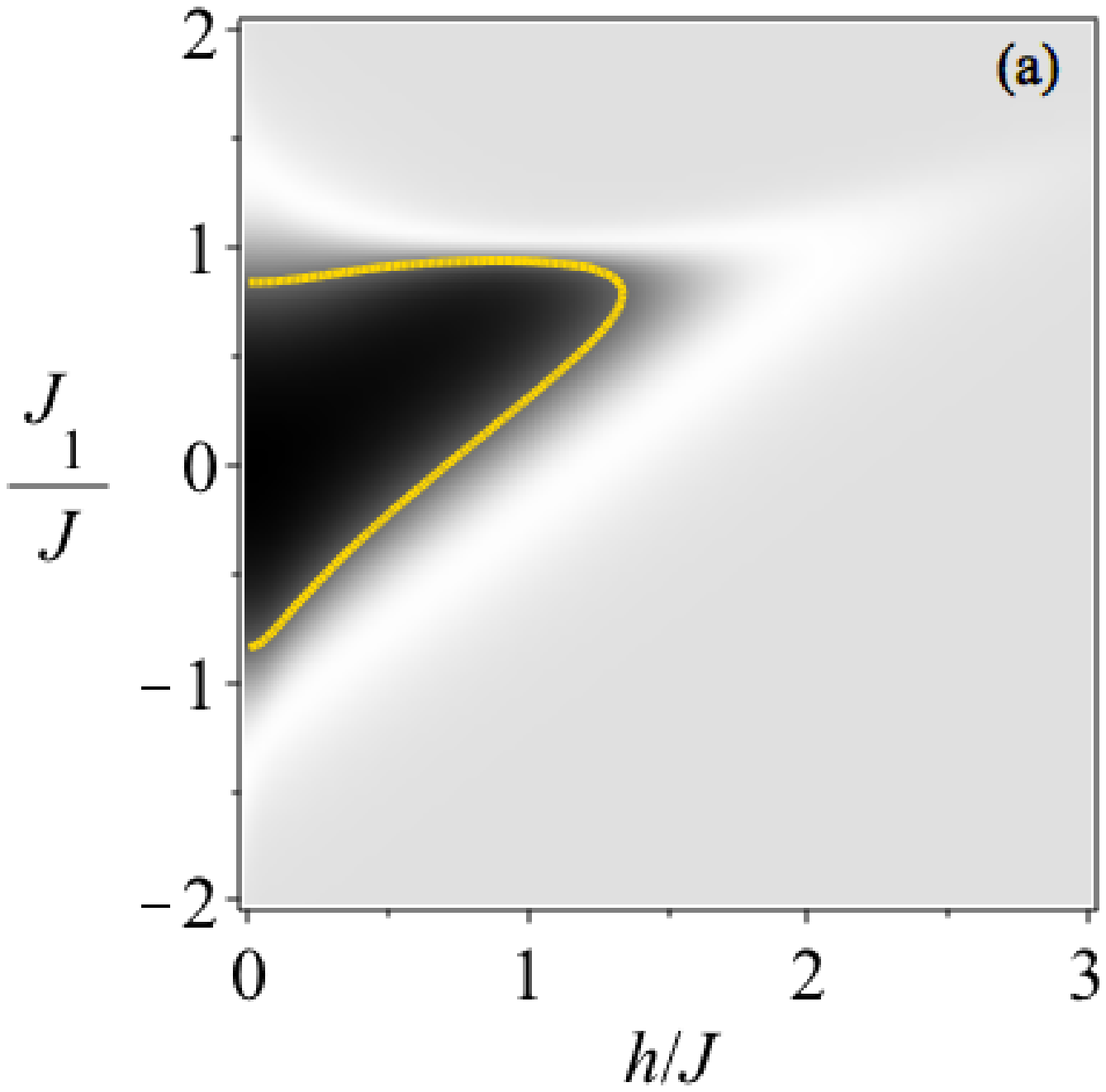}\includegraphics[scale=0.3]{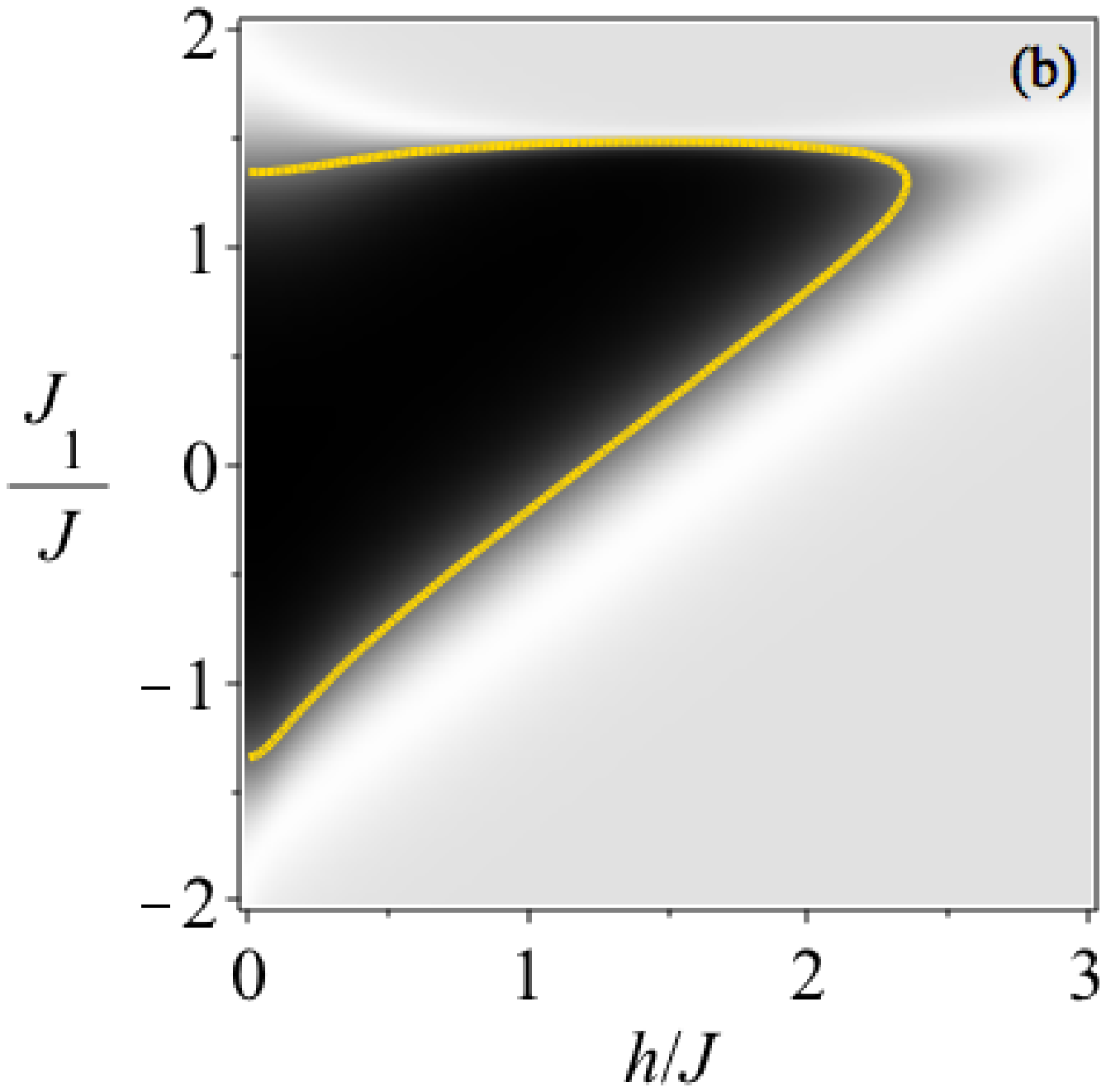}\caption{\label{fig:J1-h}Density plot of average fidelity $F_{A}$ as a function
of $h/J$ versus $J_{1}/J$ for $T/J=0.2$. (a) $\Delta=1$. (b) $\Delta=2$.
For both figures, the yellow curve is the contour for $F_{A}=2/3$.}
\end{figure}

In Fig. \ref{fig:J1-h}, we show the properties of the average fidelity
$F_{A}$ versus parameter $J_{1}/J$ and magnetic field $h/J$. From
the Fig. \ref{fig:J1-h}(a), we note that the average fidelity becomes
$F_{A}>\frac{2}{3}$ for $|\frac{J_{1}}{J}|\lesssim0.84$ and magnetic
field $h/J\lesssim1.33$. In Fig.\textcolor{magenta}{{} }\ref{fig:J1-h}
(b), it is shown that the model is acceptable for teleportation when
$|\frac{J_{1}}{J}|\lesssim1.33$ and $\frac{h}{J}\lesssim2.35$. Similar
to the previous case, the surrounding yellow curve is the boundary
between quantum teleportation would be successful or not.

\begin{figure}
\includegraphics[scale=0.23]{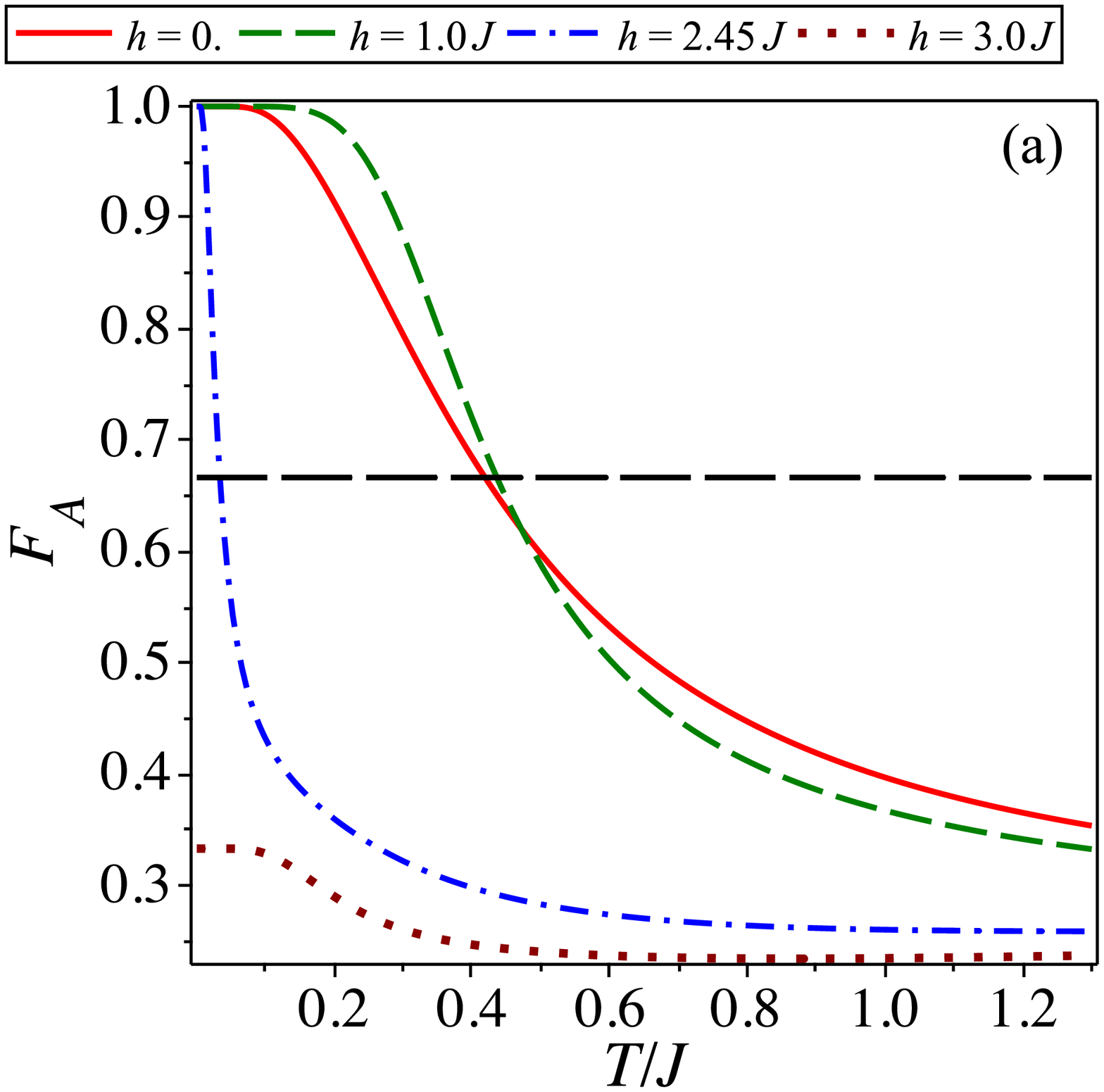}\includegraphics[scale=0.23]{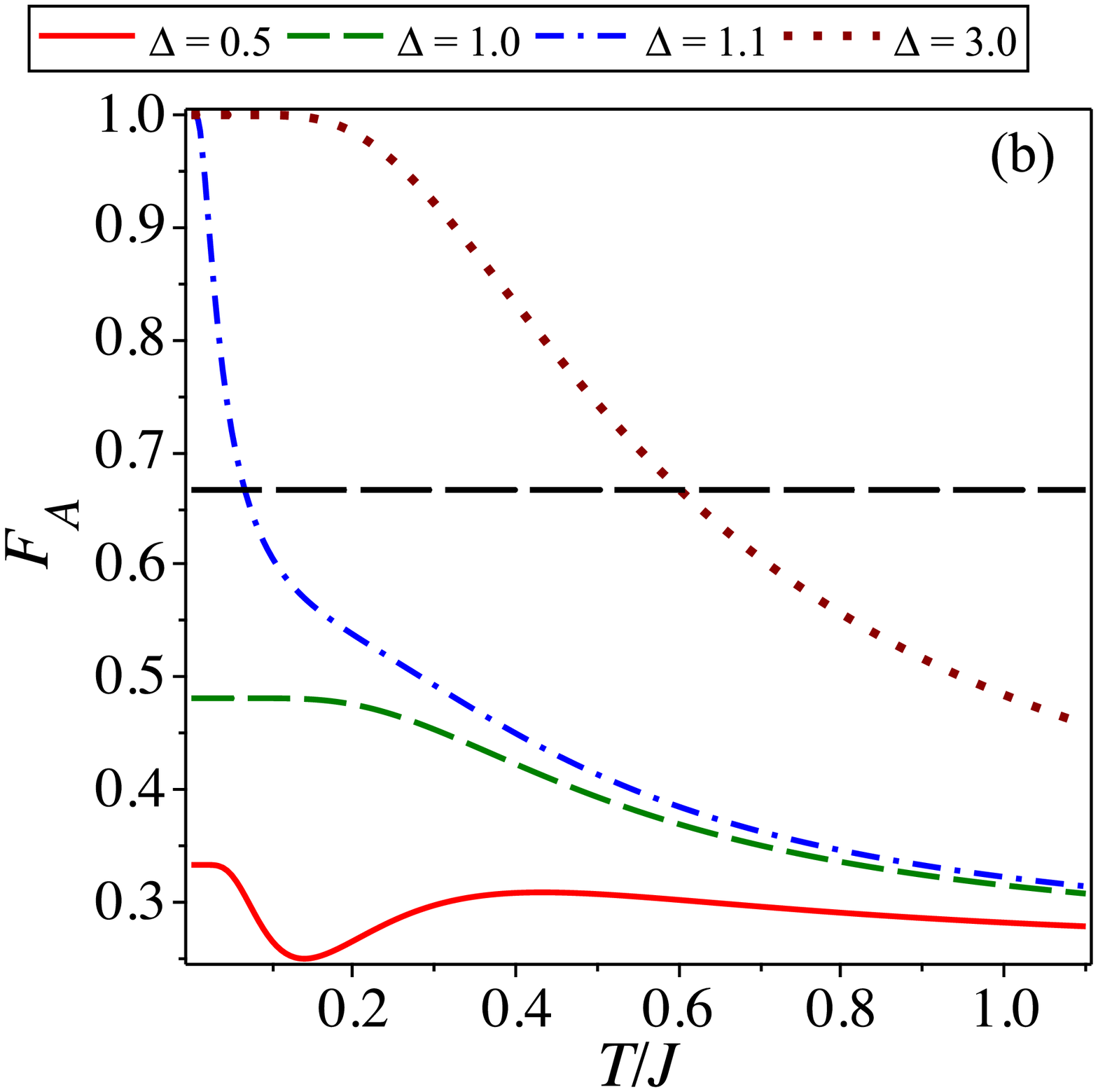}\caption{\label{fig:Fa-T}(Color online) Average fidelity $F_{A}$ is plotted
as a function of the temperature $T/J$ for $J_{1}/J=1$. (a) For
$\Delta=2$. (b) For $h/J=0$. Horizontal dashed line indicates the
2/3 constant line.}
\end{figure}

On the other hand, in Fig. \ref{fig:Fa-T} is shown the average fidelity
$F_{A}$ as a function of the temperature $T/J$ for different values
of $h/J$ and $\Delta$, assuming fixed values $J_{1}/J=1$. In the
figures the horizontal dashed lines at $F_{A}=2/3$ denote the limit
of quantum fidelities. The Fig. \ref{fig:Fa-T}(a) shows clearly that,
the magnetic field can enhance the average fidelity in interval $0\leqslant h/J<2.5$.
Furthermore, the plot also shows that when increasing the magnetic
field $h/J\geqslant2.5$, the teleportation will not succeed since
$F_{A}\leqslant2/3$. This occurs because the systems is in the interphase
between the entangled state and the unentangled state at zero temperature
(see Ref. \cite{mrojas}). Whereas, in Fig. \ref{fig:Fa-T}(b), we
can notice that the average fidelity gradually decreases below $2/3$,
when increases $\Delta$ up to $\Delta=1$. For $\Delta=1.1$, the
average fidelity leads to a fixed value $1$, in the low temperature
region, when temperature increases, the value of $F_{A}$ quickly
decays below $2/3$. With the increase of the anisotropy parameter
to $\Delta=3$, the average fidelity $F_{A}$ enhances and then decreases
monotonically as soon as the $T/J$ increase. Thus, we conclude the
quantum teleportation protocol will succeed for large $\Delta$ and
for a given magnetic field. 

\begin{figure}
\includegraphics[scale=0.23]{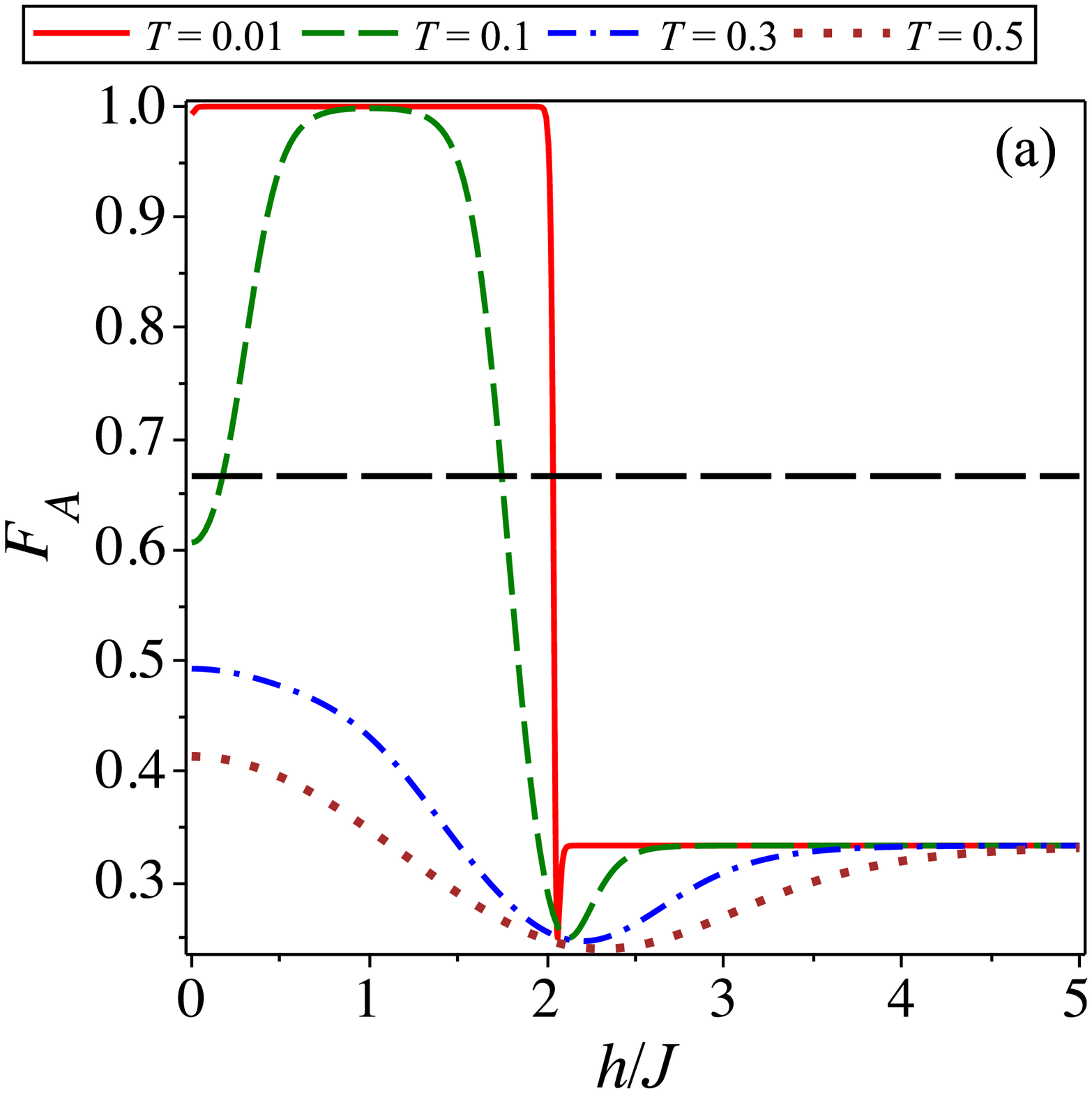}\includegraphics[scale=0.23]{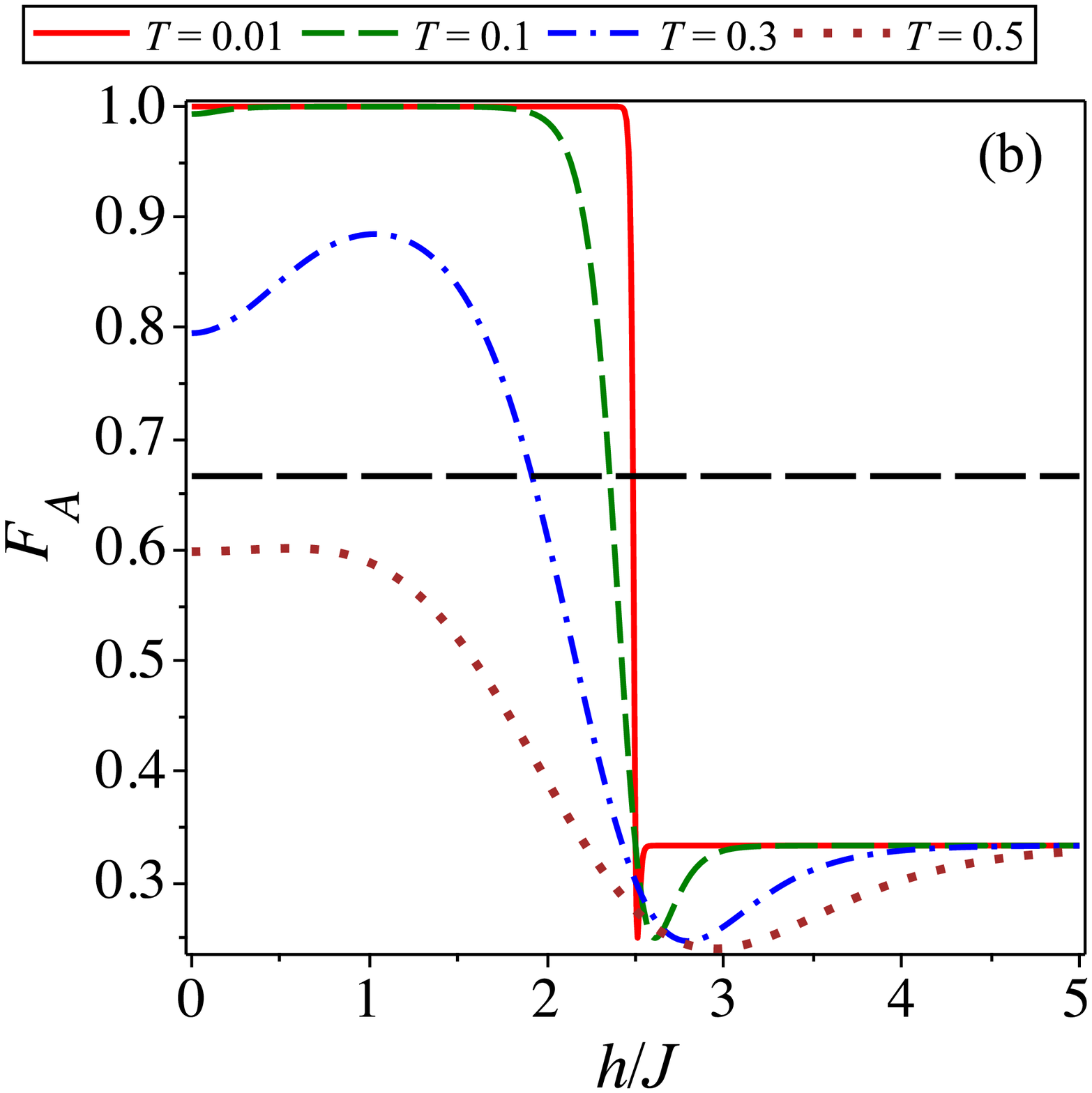}\caption{\label{fig:Fvsh}(Color online) Average fidelity $F_{A}$ is plotted
as a function of the magnetic field $h/J$ for different values of
temperature and $J_{1}/J=1$. (a) For $\Delta=1.1$. (b) For $\Delta=2$.
Horizontal dashed line indicates the 2/3 constant line.}
\end{figure}

Furthermore, to illustrate the effect of the magnetic field $h/J$
on the average fidelity, we plot $F_{A}$ as a function of $h/J$
at different temperature $T/J$. In Fig. \ref{fig:Fvsh}(a), we can
observe that for $\Delta=1.1$, the average fidelity is constant and
equal to $F_{A}=1$, for temperature close to zero $T/J=0.01$, and
then drops suddenly to $F_{A}\approx0.25$ at a critical value $h/J=2.06$.
For $h/J>2.06$, the average fidelity gains an asymptotic revitalization
up to\textcolor{red}{{} }$F_{A}\rightarrow1/3$, after reaching to its
minimum value. For temperature $T/J\gtrsim0.1$ the average fidelity
is less than $2/3$ in $h/J=0$. When the magnetic field is introduced,
the average fidelity attain the maximum and then, the average fidelity
$F_{A}$ decreases leading to $F_{A}\approx0.25$. However, there
is a revival of average fidelity reaching value\textcolor{red}{{} }$F_{A}\rightarrow1/3$. 

On the other hand, we can observe clearly that when $T/J\geq0.3$,
the average fidelity is always smaller than $2/3$. In Fig. \ref{fig:Fvsh}(b),
we plot the average fidelity as a function of the magnetic field,
for $\Delta=2$. One can find when $T/J$ is equal to $0.5$, it is
no longer possible to observe the quantum communication in absence
of magnetic field $h/J=0$. Here, the average fidelity also presents
the same interesting feature, the curves almost vanishes and recovery
the average fidelity for a particular value of the magnetic field
$h$. After achieving the minimum value, the average fidelity increases
monotonically when increasing the magnetic field up to $F_{A}\rightarrow1/3$.

\begin{figure}
\includegraphics[scale=0.23]{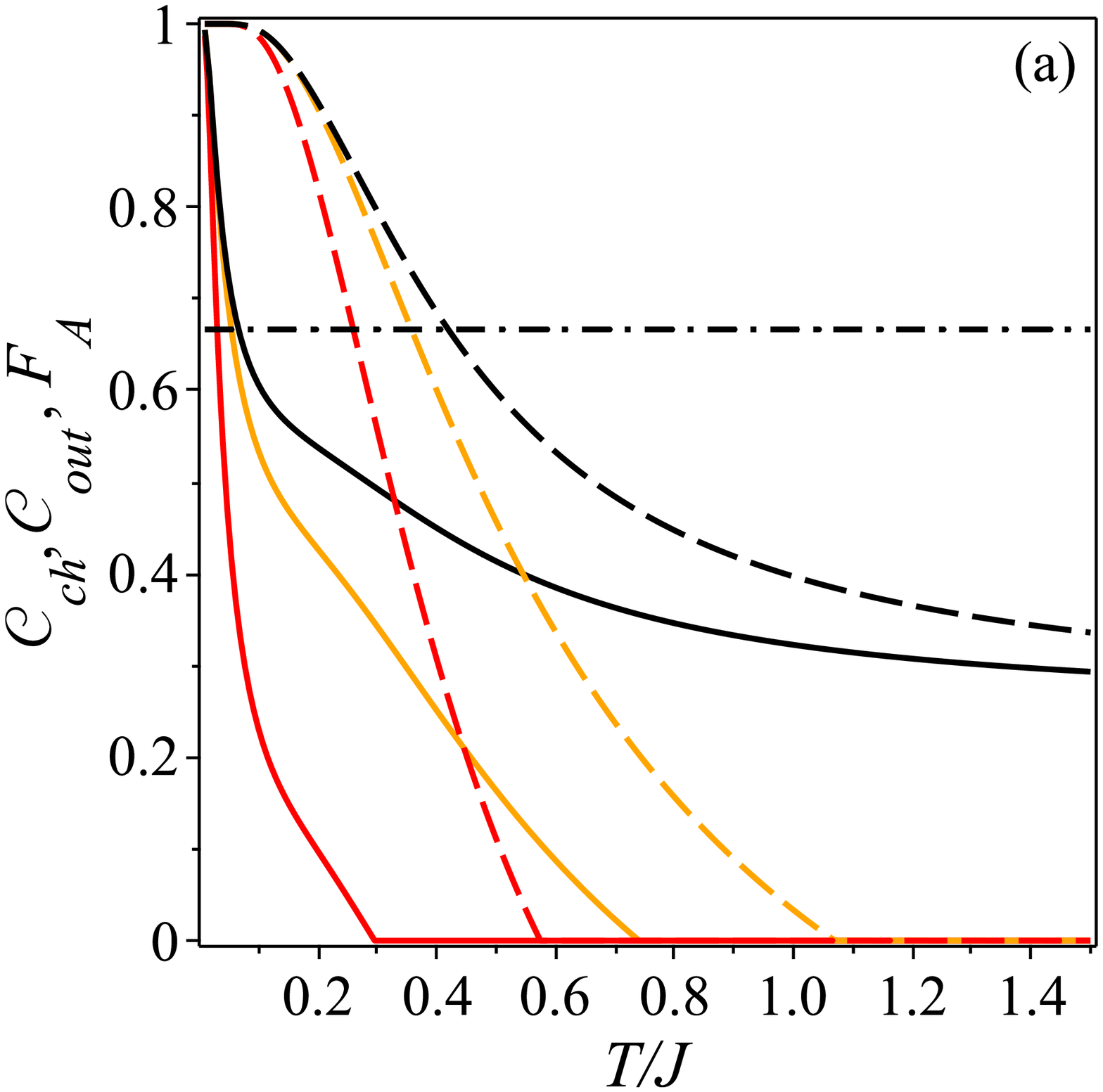}\includegraphics[scale=0.23]{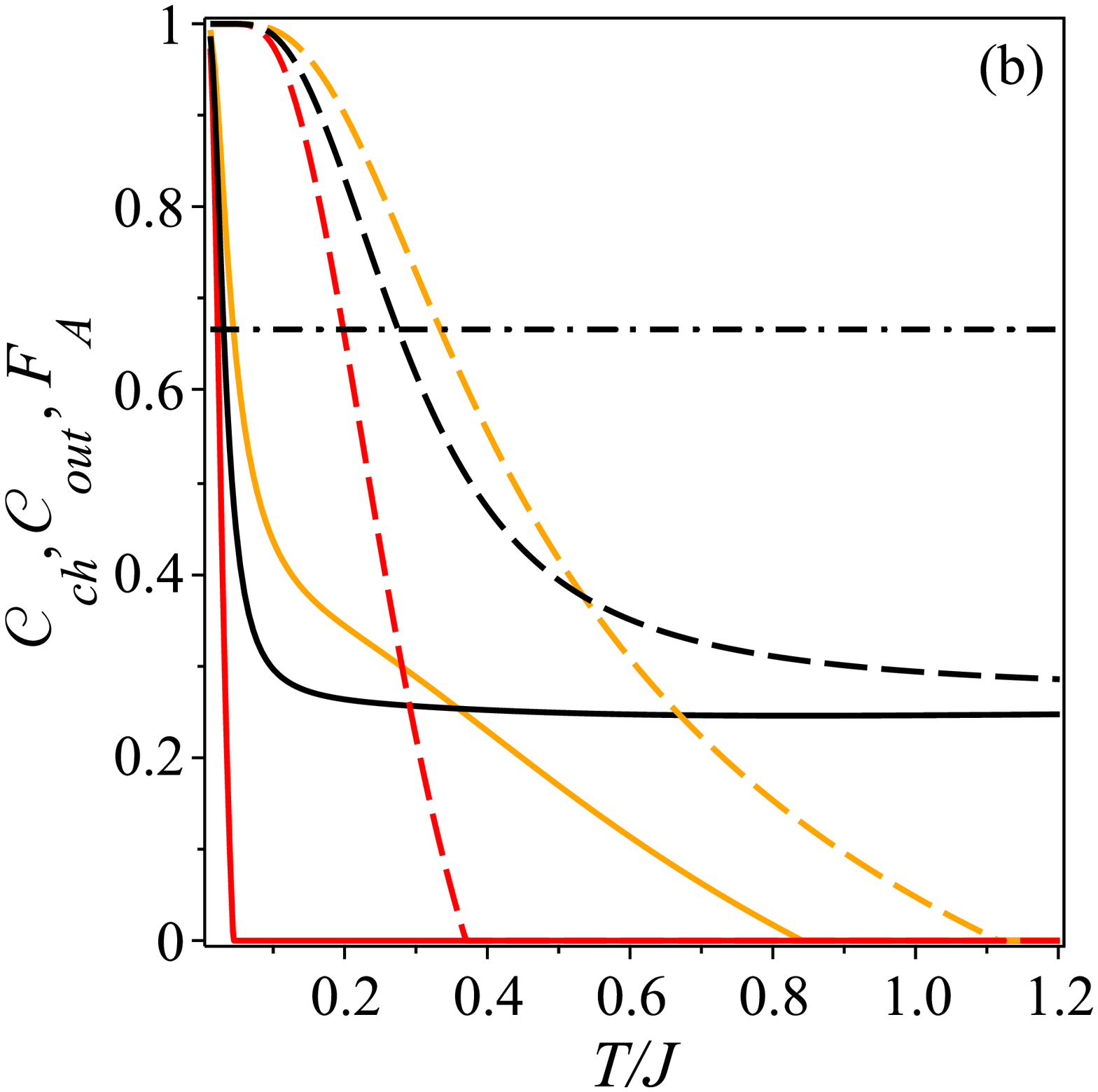}\caption{\label{fig:compar}(Color online) Comparison of quantum channel concurrence
(yellow), output concurrence (red) and average fidelity (black) versus
$T/J$ for $J_{1}/J=1$, $\Delta=1.1$ (\textit{solid line}) and $\Delta=2$
(\textit{dashed line}). (a) For $h/J=0$. (b) For $h/J=2$. Horizontal
dashed line indicates the 2/3 constant line.}
\end{figure}

Finally, to observe the effects of output concurrences and average
fidelity of quantum teleportation. We plot the quantum channel concurrence
$\text{\ensuremath{\mathcal{C}}}_{ch}$ (yellow curves), output concurrence
$\mathcal{C}_{out}$ (red curves) and average fidelity $F_{A}$ (black
curves) as a function of temperature $T/J$ for different values of
the $h/J$ and $\Delta$. In Fig. \ref{fig:compar}(a) are displayed
$\text{\ensuremath{\mathcal{C}}}_{ch}$, $\text{\ensuremath{\mathcal{C}}}_{out}$
and $F_{A}$ versus $T/J$. For null magnetic field $h/J=0$ and fixed
$\Delta=1.1$ (solid lines), the quantum channel concurrence and output
concurrence disappears in the threshold temperature $T_{th}/J\approx0.74$
and $T_{th}/J\approx0.3$ respectively. While the average fidelity
is successful up to temperature $T/J\approx0.1$, and $F_{A}$ gradually
decreases when increases  $T/J$ until reaching $F_{A}\approx0.25$.
For $\Delta=2$ (dashed line), the quantum channel concurrence $\text{\ensuremath{\mathcal{C}}}_{ch}$
vanishes for $T_{th}/J\gtrapprox1.07$. The threshold temperature
for output concurrence $\text{\ensuremath{\mathcal{C}}}_{out}$ occurs
at $T_{th}/J\gtrapprox0.58$. It is also observed that the quantum
communication is enhanced with the increase of $\Delta$ ($F_{A}>\frac{2}{3}$,
represented as dash-doted line). On the other hand, for higher temperature,
$F_{A}$ decays monotonically and it approaches to $0.25$. In Fig.
\ref{fig:compar}(b) for a fixed value of $h/J=2$, we display $\text{\ensuremath{\mathcal{C}}}_{ch}$,
$\text{\ensuremath{\mathcal{C}}}_{out}$ and $F_{A}$ as functions
of $T/J$ and fixed $\Delta=1.1$ (solid lines). One can notice that,
the quantum channel concurrence and output concurrence disappears
in the threshold temperature $T_{th}/J\approx0.84$ and $T_{th}/J\approx0.043$
respectively. While $F_{A}$ decreases monotonically until $F_{A}\approx0.25$
as soon as the temperature increases. On the other hand, for $\Delta=2$
(dashed lines) the threshold temperature for $\text{\ensuremath{\mathcal{C}}}_{ch}$
is $T_{th}/J=1.12$ and for $\text{\ensuremath{\mathcal{C}}}_{out}$
becomes $T_{th}/J=0.37$. However the average fidelity leads asymptotically
to $F_{A}\approx0.25$. 

\section{Conclusions}

In this article, we have studied the quantum teleportation of two
qubits. The quantum teleportation performing through a quantum channel
composed by a couple of Heisenberg dimers in an Ising-$XXZ$ diamond
chain structure. When the input state is a pure state, we can apply
the concept of fidelity as a useful quantity to study the of teleportation
performance by a quantum channel \cite{horo}. Assuming the teleported
qubits in an arbitrary state, we have obtained analytical results
for the quantum channel concurrence, output concurrence, and the average
fidelity. We discussed in detail the effects of coupling parameters
of the Ising-Heisenberg diamond chain (quantum channel), as a function
of magnetic field and temperature dependence for the output concurrence,
quantum channel concurrence, and average fidelity. Displaying the
results as a function of the anisotropy coupling $\Delta$ and external
magnetic field $h$. Therefore, we observe the magnetic field $h$
and the anisotropy parameter $\Delta$ influence strongly in the output
concurrence and average fidelity. Thus, when the magnetic field increased,
the output concurrence is stimulated favoring the teleportation success
for a larger region. However, for a sufficiently strong magnetic field,
the output concurrence gradually decreases, and the entangled state
teleportation will not succeed anymore. In figures \ref{fig:TvsJ1}
and \ref{fig:J1-h} have illustrated this effect, the darkest region
contoured by a solid line shows the region in which the quantum teleportation
could well succeed, whereas for the outside of this curves means the
quantum teleportation would fail. 

On the other hand, we observe the quantum teleportation increases
when increases of anisotropy parameter $\Delta$, then we conclude
the parameter $\Delta$ is the more efficient control parameter of
quantum communications illustrated in figures \ref{fig:Fvsh} and
\ref{fig:compar}.

\section*{Acknowledgment}

O. Rojas, M. Rojas and S. M. de Souza thank CNPq, Capes and FAPEMIG
for partial financial support. O. R. also thanks ICTP for financial
support and hospitality.

\end{document}